\documentclass[12pt]{article}

\usepackage[pdfpagelabels]{hyperref}

\usepackage[usenames,dvipsnames]{color}
\usepackage{graphicx,epsfig,xcolor}
\usepackage{mathtools}

\hypersetup{colorlinks=true, linkcolor=violet, urlcolor=blue, citecolor=blue}

\usepackage{cite}
\usepackage[english]{babel}
\usepackage{amssymb,amsfonts,amsmath}
\usepackage{verbatim}
\usepackage{bbm,bbold,bm}
\usepackage{tensor}
\usepackage{slashed}
\usepackage{braket}
\usepackage{geometry}
\numberwithin{equation}{section} 
\usepackage{physics}

\title{\bf Scale without conformal symmetry in hydrodynamics}

\author{\small Evangelos Afxonidis$^{a,}$\footnote{evangelos.afxonidis@uniovi.es} , Jewel Kumar Ghosh$^{b,c,}$\footnote{jewel.ghosh@iub.edu.bd} , Daniele Musso$^{a,}$%
\footnote{mussodaniele@uniovi.es} ,\\ \small
Daniel Naegels$^{d,}$%
\footnote{dn441@cam.ac.uk}, Ignacio Salazar Landea$^{e,}$\footnote{peznacho@gmail.com}
}

\date{}

\begin{document}

\maketitle
\vspace{-22pt}
\begin{center}\it{\small 
$^a$Department of Physics and Instituto de Ciencias y Tecnolog\'{\i}as Espaciales de Asturias (ICTEA) Universidad de Oviedo, c/ Federico García Lorca 18, ES-33007 Oviedo, Spain
\\
$^b$Department of Physical Sciences, Independent University, Bangladesh (IUB), Bashundhara RA, Dhaka 1229, Bangladesh \\
$^c$Center for Computational and Data Sciences (CCDS), Independent University, Bangladesh, Dhaka 1229, Bangladesh \\
$^d$Department of Applied Mathematics and Theoretical Physics, University of Cambridge, Wilberforce Road, Cambridge, CB3 0WA, United Kingdom \\
$^e$Instituto de F\'\i sica de La Plata - CONICET, C.C. 67, 1900 La Plata, Argentina \\}
\end{center}
\vspace{15pt}
\begin{abstract}
Scale without conformal symmetry corresponds to an inhomogeneous conservation equation for the virial current sourced by the trace of the energy-momentum tensor. Fluids that are just scale-invariant differ qualitatively from their conformal counterparts, and generic dissipation effects relax the hydrodynamic response over sufficiently long time scales. Remarkably, this holds true already at the ideal order. Spontaneously broken scale symmetry does not, in general, add any new mode to the hydrodynamic sector.
\end{abstract}

\vspace{1.5cm}
\noindent
{\footnotesize \textit{Dedicated to the innocent lives lost in the tragic plane crash on Milestone School and College in Dhaka on 21 July 2025.}}

\tableofcontents

\newpage

\section{Introduction and main results}
\label{sec:intro}

Since Weyl's attempts to unify electromagnetism and gravity relying upon scale symmetry \cite{Weyl:439957}, scale-invariant systems have attracted significant attention in physics.
But scale symmetry is known to be subtle. As a spacetime symmetry, dilatation breaking is not controlled by the original formulation of Goldstone theorem and, differently from rigid spacetime translations, rigid scale transformations affect the spacetime metric.
Important fundamental open questions regarding naturalness problems in cosmology and particle physics are related to scale symmetry and its breaking \cite{Coradeschi:2013gda}. Whether scale symmetry is necessarily enhanced to conformal symmetry in quantum field theory is by itself an aspect that motivated an extensive and intense research effort, whose results are not yet completely systematized \cite{Nakayama:2013is}.

Scale symmetry enhanced to conformal symmetry has been extremely significant also in the context of hydrodynamics for at least two reasons, one experimental and the other theoretical. The former being the relevance of conformal hydrodynamics to describe the quark gluon plasma \cite{Romatschke:2017ejr,Busza:2018rrf}, the latter arising from the importance of conformal hydrodynamics in accounting for the low-energy regime of holographic systems \cite{Son:2007vk}. 

Maybe because of the practical success of conformal hydrodynamics, some aspects of the role of scale symmetry in hydrodynamics have not been discussed thoroughly. The first is the consequence of having scale symmetry without conformal enhancement. The second is a discussion of how scale symmetry is spontaneously broken and the consequences of this breaking.

Hydrodynamics relies on conservation equations and constitutive relations for the conserved currents \cite{Eckart:1940te,landau1987fluid,Kovtun:2012rj}. In the modern approach, both are derived via functional methods \cite{Jensen:2012jh,Kovtun:2019hdm}. In the presence of Nambu-Goldstone bosons, the linear hydrodynamic system of equations typically needs to be supplemented by Josephson relations that directly descend from the definition of the chemical potentials for the spontaneously broken symmetries, the paradigmatic example being the $U(1)$ superfluid. To some extent, the dilaton constitutes an exception to this. Specifically, the need of an extra Josephson equation for broken scale symmetry depends on whether one is willing to describe a condensate whose fluctuations are parametrized by the dilaton. The peculiarity of the dilaton arises from the fact that, even in the absence of such a condensate, the dilaton does not decouple completely from the low-energy description. Clearly this relates to the fact that assuming a finite temperature as one does in hydrodynamics already implies the breaking of scale symmetry. This is to be contrasted to the normal phase of the standard $U(1)$ superfluid where the symmetry is unbroken and the Nambu-Goldstone mode absent from the hydrodynamic description. As an outcome of our analysis, we even show that the presence of a dilatonic condensate leads to an inconsistent thermodynamics, at least in the simplest neutral hydrodynamic systems. Our main results are the following.
\begin{enumerate}
    \item 
    Scale without conformal symmetry leads to generic dissipation effects in all the hydrodynamic modes. This implies that, over long enough time scales, the hydrodynamic response relaxes completely.
    \item 
    The sector associated with the spontaneous breaking of scale symmetry does not, in general, contribute new hydrodynamic modes.
\end{enumerate}

Especially the former result can to some extent be intuitively expected a priori. The equation describing scale conservation in the non-conformal case equates the divergence of the virial current and the trace of the energy-momentum tensor. It thus takes the mathematical form of an inhomogeneous equation for the non-conservation of a current. This is the technical origin of the imaginary terms that leads to dissipation in the low-energy hydrodynamic modes. Since all fields enter the energy-momentum tensor, these dissipation effects affect all hydrodynamic modes. Note that such dissipation enters already at the so-called ideal order, that is, zero order in derivatives within the constitutive relations.

The present study also has some technically interesting by-products. We derived the scaling Ward-Takahashi identity with background field methods within the functional formalism and confirmed the identification of the Weyl and virial currents \cite{Zanusso:2023vkn}. We checked this result explicitly against known examples of scale symmetric but non-conformal systems \cite{El_Showk_2011,Riva_2005} (see Appendix \ref{sec:sio}). As another interesting result, we showed that asking for the closure of the gauge transformations for the background fields imposes the scaling dimension of the Weyl connection to be null (see Appendix \ref{app:closing_the_algebra}). Eventually, we comply with and thus confirm some existing results previously obtained in the Weyl-covariant formulation of hydrodynamics \cite{Bhattacharyya:2008mz,Loganayagam:2008is} and in the applications of conformal hydrodynamics \cite{Herzog:2008he}.

The structure of the paper goes as follows. In Section \ref{sec:WT} we derive the Ward-Takahashi identities encoding the conservation laws by means of self-consistent background field methods (details are provided in Appendix \ref{app:Ward-Takahashi} and Appendix \ref{app:closing_the_algebra}). Section \ref{sec:hydro} is dedicated to the derivation of the constitutive relations and the full system of linear hydrodynamic equations. The mode solutions of the linear hydrodynamic system are studied in Section \ref{sec:modes}. Eventually, we gather some final comments and sketch the future prospect in Section \ref{sec:remfut}.

\section{Ward-Takahashi identities}
\label{sec:WT}

We derive the Ward-Takahashi identities for the global symmetries of the system, namely spacetime translations, spatial rotations, Lorentz boosts and scale transformations, by means of functional methods. The global scale symmetry is implemented \emph{à la} Weyl \cite{Iorio:1996ad}: the fields and the vielbeins $e_\mu^a$ transform, while the coordinates and derivatives are invariant. All the appearances of the metric must be explicit, in particular in the integration measure $d^dx\,\sqrt{-g}$, where $g_{\mu\nu} = e^a_\mu e^b_\nu \eta_{ab}$.

The system is coupled to a connection $\Omega_\mu$ for scale transformations. This and the metric $g_{\mu\nu}$ are non-dynamical background fields whose variations define the currents through
\begin{equation}
    \label{eq:def_cur}
    \delta \Gamma 
    \equiv 
    \int d^dx \,
    \sqrt{-g} \left(
    -\frac{1}{2} T^{\mu\nu} \delta g_{\mu\nu}
    - I^\mu \delta \Omega_\mu
    \right)\ .
\end{equation}
To derive the Ward-Takahashi identities, we employ the Noether procedure, which involves promoting global symmetries to local ones and requiring the theory to remain invariant under these local transformations. The Ward-Takahashi identities for boosts and rotations correspond to the stress energy-momentum tensor being symmetric $T^{\mu\nu}=T^{\nu\mu}$. In the functional framework, this is an automatic consequence of the symmetry of the metric. The variations of the background fields due to the local version of the translation and dilatation symmetries are respectively:
\begin{align}
    \label{eq:gauge_transformations}
    \delta g_{\mu\nu} = 
    {\cal L}_\xi g_{\mu\nu}
    - 
    2 \sigma g_{\mu\nu}
    \ , \qquad
    \delta \Omega_\mu = 
    {\cal L}_\xi \Omega_\mu 
    + \partial_\mu \sigma
    \ ,
\end{align}
where $\xi^\mu$ parametrizes the diffeomorphisms, $\sigma$ parametrizes scale transformations, and $\mathcal L_\xi$ denotes the Lie derivative with respect to $\xi^\mu$. Note that, while the metric with lower indices has scaling dimension $-2$, the connection $\Omega_\mu$ has trivial scaling dimension.%
\footnote{The closure of the background gauge transformation algebra imposes the vanishing of the scaling dimension for the connection $\Omega_\mu$, see Appendix \ref{app:closing_the_algebra}. A similar argument for Galilean cases is studied in \cite{Jensen:2014aia}.} 
In order to get the Ward-Takahashi identities
\begin{align}
    \label{eq:WTI}
    \partial_\mu T^{\mu\nu} = 0
    \ , \qquad 
    \partial_\mu I^\mu = T^\mu_{\ \mu}\ ,
\end{align}
we perform the gauge variation for each symmetry parameter and take into account the necessary integrations by parts. Note that in \eqref{eq:WTI} we have also specified the analysis to the background where%
\footnote{Further details on the Ward-Takahashi identities, also in the presence of non-trivial backgrounds, are given in Appendix \ref{app:Ward-Takahashi}.}
\begin{equation}
    g_{\mu\nu} = \eta_{\mu\nu}
    \ , \qquad 
    \Omega_\mu = 0\ .
\end{equation}
We refer to $I^\mu$ as the Weyl current. We underline that the above derivation shows that the Weyl current coincides with the virial current.%
\footnote{We confirm this point and also the structure of the scaling Ward-Takahashi identity against some explicit examples of scale-invariant-only models in Appendix \ref{sec:sio}. See \cite{Zanusso:2023vkn} for a related discussion.} We do not consider the possible presence of anomalies since they would not affect the hydrodynamic analysis at ideal order \cite{Baier:2007ix}. The scale Ward-Takahashi identity can be cast as the conservation of a scaling current
\begin{equation}
    \label{eq:scaling_conservation_Weyl}
    J^{\mu}_D \equiv I^\mu - x^\nu T^\mu_{\ \nu}\ .
\end{equation}

Considering also special conformal transformations would additionally imply $T^\mu_{\ \mu} = 0$ (see Appendix \ref{app:conformal}). In the conformal case, the virial term can be absorbed into the energy-momentum tensor through an improvement \cite{Wess1960,Polchinski:1987dy}. 
If we insist that the shape of the scaling Ward-Identity is $\partial_\mu I^\mu = T^\mu_{\ \mu}$, then it means that we are in the scale-invariant but not conformal case.%
\footnote{
\hypertarget{foot:d2}{} \hspace{-4pt}This point is discussed in detail in \cite{Dymarsky:2013pqa}. They also mention the possible existence of specific cases where the presence of operators whose generalized dimension is $d-2$ could make it impossible to improve the energy-momentum tensor in such a way that in \eqref{eq:def_cur} there are no extra terms. Such extra terms would nonetheless add up to those we already consider. Thus, it is reasonable to expect that the dissipation phenomena that we describe (due to a derivative mismatch among the two sides of the scaling Ward-Takahashi identity \eqref{eq:WTI}) would remain. Let us also remind the reader that, for a conformal theory in $d=2$, the energy-momentum tensor is automatically traceless, without any improvement ambiguity \cite{Dymarsky:2013pqa}. This also implies that our approach does not apply to $d=2$, where we cannot assume a (unitary) scale invariant but non-conformal system.}

The triviality of the scaling dimension for the connection $\Omega_\mu$ implies that the scaling dimension for the Weyl current coincides with the spacetime dimension $d$. The physical dimension of the energy-momentum tensor $T_\mu^{\ \nu}$ is that of a spatial density of energy, namely $d$. Recalling that the derivative is dimensionless, the scaling Ward-Takahashi identity is dimensionally consistent. Note also that the dimension of $T^{\mu\nu}$ is $d+2$ \cite{Gubser:2010ui}.

\section{Hydrodynamic equations}
\label{sec:hydro}

In this section we perform a linear hydrodynamic analysis of scale-invariant, ideal, neutral fluids. We first derive the constitutive relations from a generating functional consistently coupled with a background and then obtain the linear hydrodynamic equations enforcing the Ward-Takahashi identities and, depending on the physical situation, a Josephson relation for the dilaton.

\subsection{Constitutive relations}

The thermodynamic equilibrium is geometrically encoded in a time-like Killing vector $k^\mu$ which defines the local temperature and local fluid frame through \cite{Jensen:2012jh}
\begin{align}
    \label{eq:Tdef}
    T &= \frac{1}{\sqrt{k^2}}
    \ , \qquad 
    u^\mu = \frac{k^\mu}{\sqrt{k^2}}
    \ .
\end{align}
Both the fluid frame 4-velocity $u^\mu$ and the temperature $T$ are here dimensionless. The contraction of the Killing vector $k^2 = k^\mu k^\nu \hat g_{\mu\nu}$ is made with respect to the dimensionless metric $\hat g_{\mu\nu} \equiv e^{2\tau} g_{\mu\nu}$, where the dilaton field $\tau$ shifts under scalings (some details about the introduction of $\tau$ in the computation are given in Appendix \ref{app:Ward-Takahashi}). Note that the normalization of the 4-velocity reads $u^\mu u^\nu \hat g_{\mu\nu} = 1$.

We can define the ``supercurrent" $V_\mu$, analogous to that of the standard $U(1)$ superfluid, 
\begin{eqnarray}
    \label{eq:def_V}
    V_\mu \equiv  \partial_\mu  \tau  - \Omega_\mu\ .
\end{eqnarray}
The other thermodynamic scalar quantities that can be built using the 4-velocity $u^\mu$ and the supercurrent $V^\mu$ correspond to an analog chemical potential for the Weyl symmetry, and a scalar related to the contraction of $V^\mu$ with itself:
\begin{align}
    \label{eq:mudef}
    \nu \equiv u^\mu V_\mu
    \ , \qquad 
    W 
    \equiv 
    \frac{1}{2} V_\mu V^\mu
    =
    \frac{1}{2} \hat g^{\mu\nu} V_\mu V_\nu
    \ .
\end{align}
The supercurrent, as well as the scalars defined above, are dimensionless. The inverse metric $\hat g^{\mu\nu} \equiv e^{-2\tau} g^{\mu\nu}$ is dimensionless too and does not scale. 

The starting assumption for the hydrodynamical study at ideal order identifies the effective Lagrangian with (minus) the local Landau grand potential, whose density coincides with the pressure,%
\footnote{
Working at ideal order (\emph{i.e.} at zero order in gradients in the action) means that we exclude standard dissipation effects, which enters at linear order, see for example \cite{Kovtun:2012rj}. This means also that we are not in a position to discuss constraints coming from entropy production \cite{Bhattacharyya:2013lha}.}
\begin{equation}
\label{eq:effac}
 \Gamma 
 = \int d^dx\, \sqrt{-g}\, {\cal L}_{\text{eff}}
 = \int d^dx\, \sqrt{-g} \, e^{d\tau}\left[P + {\cal O}(\partial )\right]\ .
\end{equation}
The variation of the dimensionless pressure $P$ is related to the variations of the dimensionless thermodynamic variables (the scalars defined above) through standard thermodynamic relations
\begin{align}
    dP
    &=
    \left(\frac{\partial P}{\partial T}\right)_{\nu,W} dT
    +
    \left(\frac{\partial P}{\partial \nu}\right)_{T,W} d\nu
    +
    \left(\frac{\partial P}{\partial W}\right)_{T,\nu} dW
    \ .
\end{align}
For later convenience we denote the thermodynamic derivatives above as follows
\begin{equation}
    \label{eq:var_pre}
    dP 
    = 
    s\, dT 
    + n\, d\nu 
    + h^2 dW 
    \ ,
\end{equation}
where $h$, by analogy with the $U(1)$ superfluid, is associated to the presence of a condensate. 

The equations we obtained in \eqref{eq:Tdef} and \eqref{eq:mudef} relate thermodynamic variables and field theoretical quantities (these latter are background and Nambu-Goldstone fields). We can therefore make this connection explicit at the level of linear variations,
\begin{align}
    \label{eq:var1}
    \delta T &= -\frac{1}{2} T u^\alpha u^\beta e^{2\tau} \delta g_{\alpha\beta}
    -T \delta\,\tau
    \ , \qquad 
    \delta u^\mu = -\frac{1}{2} u^\mu u^\alpha u^\beta e^{2\tau} \delta g_{\alpha\beta}
    -u^\mu \delta\tau
    \ , \\ \label{eq:var2}
    \delta \nu &= 
    - \frac{1}{2} \nu\, u^\alpha u^\beta e^{2\tau} \delta g_{\alpha\beta}
    - u^\mu \delta
    \Omega_\mu 
    - \nu\, \delta \tau
    +  u^\mu\partial_\mu \delta\tau
    \ , \\ \label{eq:var3}
    \delta W &= 
    -\frac{1}{2} V^\mu V^\nu e^{2\tau} \delta g_{\mu\nu}
    - V^\mu \delta\Omega_\mu
    -2W \delta\tau
    + V^\mu \partial_\mu \delta \tau
    \ .
\end{align}

We are now able to derive the constitutive relations for the currents,
\begin{align}
\label{eq:Tmunu}
    T^{\mu\nu} &= 
    e^{(d+2)\tau} \left(-\hat g^{\mu\nu} P 
    + s\, T u^\mu u^\nu
    + n\, \nu\, u^\mu u^\nu
    + h^2 V^\mu V^\nu 
    \right)
     \ , \\ 
     \label{eq:weycur}
    I^\mu &= 
    e^{d\tau} 
    \left(
    n u^\mu 
    + h^2 V^\mu 
    \right)
    \ .
\end{align}
From \eqref{eq:weycur}, a two-fluid picture emerges. We interpret the superfluid part proportional to $h^2$ as due to a \emph{dilatonic condensate}, namely a condensate whose fluctuations are parametrized by the dilaton. There is however an important difference between the present case and the standard $U(1)$ superfluid. In the latter a limit of zero condensate completely decouples the $U(1)$ Nambu-Goldstone mode. Here, instead, if we force the condensate to vanish ($h\to 0$), the dilaton does not decouple. In fact, the variations \eqref{eq:var1} and \eqref{eq:var2} still contain terms in $\delta\tau$. As soon as there is a non-trivial energy density, scalings are spontaneously broken. In all cases, with or without $h$, the hydrodynamic equation of motion for $\delta\tau$ returns the scaling conservation equation $\partial_\mu I^\mu = T^\mu_{\ \mu}$.%
\footnote{This harmonizes with the standard literature on conformal hydrodynamics where no dilatonic condensate is taken into account. In fact, the computations in the main text make contact with the conformal case for $I^\mu = 0$. We emphasize however that, in the conformal case, the dilaton equation of motion simply reduces to the (non-dynamical) traceless condition for the energy-momentum tensor.}

We choose the following equilibrium configuration \begin{equation}
    \label{eq:Teq}
    T = \overline T + \delta T 
    \ , \qquad 
    u^\mu = \left(1,\delta \boldsymbol{v}\right)
    \ , \qquad 
    \nu = \delta \nu
    \ , \qquad 
    \tau = \bar\tau+\delta \tau 
    \ , 
\end{equation}
and
\begin{equation}
    \label{eq:all_ord}
    \Omega_\mu = 0
    \ , \qquad 
    g_{\mu\nu} = \eta_{\mu\nu }
    \ ,
\end{equation}
where the equilibrium quantities are denoted with a bar and are spacetime constant. Since $W$ is quadratic in $V^\mu$ and this latter vanishes in the background specified by \eqref{eq:Teq} and \eqref{eq:all_ord}, the fluctuations of $h^2$ decouple from the subsequent analysis. Although the Weyl chemical potential $\nu$ is allowed to fluctuate, we set its equilibrium value to zero, $\bar \nu = 0$.%
\footnote{
Recalling the definitions for the supercurrent $V_\mu$ \eqref{eq:def_V} and chemical potential $\nu$ \eqref{eq:mudef}, having $\bar\nu = 0$ is consistent with $\Omega_\mu = 0$ and a constant $\bar\tau$.
}
We emphasize that \eqref{eq:all_ord} sets the background Weyl gauge connection to zero and the geometry to be flat, these conditions hold at all orders in the hydrodynamic expansion.

\subsection{Linear fluctuations}

In this section we derive the system of hydrodynamic equations that govern the linear fluctuations. Let us start from the Josephson relation for the Weyl supercurrent. At linear order in the fluctuations, we have
\begin{equation}
    \label{eq:Vexp}
    \delta V_\mu = 
    \partial_\mu \delta \tau 
    \ .
\end{equation}
Furthermore, from the variation of $u^\mu V_\mu = \nu$, we get $\delta V_0 = \partial_0\delta\tau 
= \delta\nu$, which, after a spatial derivation, gives
\begin{equation}
    \partial_0 \delta V_i   = \partial_i \delta \nu 
    \ .
\end{equation}

At order zero, we are in the fluid frame where $u^\mu = \delta^\mu_0$. Dropping the bar over equilibrium variables, we have 
$V_\mu = \nu\, \delta_\mu^0 + V_i \delta^i_\mu$. Accordingly, the variation of the dimensionless pressure can be expressed as
\begin{align}
    \delta P 
    &= 
    s\, \delta T 
    + \left(n+h^2\nu
    \right)\delta\nu
    - h^2 V_i \delta V_i
    \ .
\end{align}
This corresponds to 
\begin{align}
    \frac{\partial P}{\partial T} = s\ ,
    \qquad
    \frac{\partial P}{\partial\nu} = n + h^2 \nu 
    \ , \qquad
    \frac{\delta P}{\partial V_i}
    = 
    -h^2 V_i 
    \ ,
\end{align}
where all the variables, except the one with respect to which we derive, are held fixed.

The derivation of the hydrodynamic conservation equations is described in Appendix \ref{app:hydrocon}. It is convenient to define the following differential operator 
\begin{eqnarray}
    \label{eq:D_operator}
    \mathcal D \equiv
    T \frac{\partial}{\partial T}
    \ .
\end{eqnarray}
Eventually, the system of hydrodynamic equations is given by
\begin{itemize}
    \item Josephson relation:
    \begin{align}
    \label{eq:hydro_joseph_tau}
    &
    0 = 
    \partial_0 \partial_i \delta \tau - \partial_i \delta \nu
    \ , 
\end{align}
We are using $\partial_i\delta\tau$ instead of $V_i$ because the system of hydrodynamic equations as a whole does not allow to be expressed in terms of the Weyl supercurrent $V_i$.
\item Scaling current conservation
    \begin{align}
    \label{eq:ska_cur}
    &
    -
    h^2\partial_i\delta V_i
    =\gamma \Bigg\{
    \delta \tau\,d\Big[s\, T
    -(d-2) P
    \Big]  
    + \delta T \left[ 
    \mathcal D
    -(d-1) 
    \right]s 
\Bigg\}\ ,
\end{align}
    \item Energy conservation
    \begin{align}
    0
    &=
    \left[ 
     d\,
     s\, T
    -(d-2) P
    \right] 
    \partial_0 \delta \tau
    + 
    \partial_0\delta T \, \mathcal D s  
    - sT \,\partial_i\delta v_i
    \ ,
\end{align}
    \item Momentum conservation
    \begin{align}
    \label{eq:hydro_mom_cons}
    0 &=
    (d-2)\, P\, \partial_i \delta\tau 
    -
    s\,T\, \partial_0\delta v_i
    + s\, \partial_i\delta T 
    \ .
\end{align}
\end{itemize}
Note that $s\ T=w$ is the enthalpy density. We have again omitted bars on equilibrium quantities to avoid notational clutter. Besides, we have considered a vanishing equilibrium chemical potential $\nu$, as discussed above. Importantly, we have also assumed that setting the Weyl chemical potential $\nu$ to zero leads to a vanishing Weyl charge density $n$.
The term within the curly brackets in the right-hand side of \eqref{eq:ska_cur} corresponds to the trace of the energy-momentum tensor \eqref{eq:antrace}; we have introduced an artificial coefficient $\gamma$ so that we can trace how the trace of the energy-momentum tensor enters the computations. This corresponds to having the following modified scale Ward-Takahashi identity $\partial_\mu I^\mu = \gamma\, T^\mu_{\ \mu}$, where $\gamma\to\infty$ connects to a conformal limit in which the Ward-Takahashi identity reduces to $T^\mu_{\ \mu}=0$.

\section{Hydrodynamic modes}
\label{sec:modes}

We study the system of hydrodynamic equations (\ref{eq:hydro_joseph_tau}$-$\ref{eq:hydro_mom_cons}) for a single Fourier mode. The harmonic fluctuation for the generic field $\psi$ takes the form 
\begin{equation}
    \label{eq:convfou}
    \delta \psi(t) = \psi_0\, e^{i \omega t - i \boldsymbol k \cdot \boldsymbol x}\ ,
\end{equation}
where $\boldsymbol k \cdot \boldsymbol x = k_i x_i$ with $i=1,...,d-1$.
We also adopt the Jacobian notation
\begin{align}
    \frac{\partial(s,n)}{\partial(T,\nu)}
     = \left|
    \begin{array}{ccc}
        \frac{\partial s}{\partial T}
        &
        \frac{\partial n}{\partial T}
        \\
        \frac{\partial s}{\partial \nu}
        & 
        \frac{\partial n}{\partial\nu}
    \end{array}\right|\ .
\end{align}
This Jacobian corresponds to the change of variables $(s,n) \rightarrow (T,\nu)$.

\subsection{Without dilatonic condensate}
\label{subsec:normal2}

In the present subsection we consider a scale-invariant but not conformal (normal) fluid. Technically, we simply add the scaling conservation equation to the system of hydrodynamic equations for a standard neutral fluid. We thus have an extra mode. In fact, on top of the sound branch, we find a purely dissipative mode with imaginary gap,
\begin{align}
    \label{eq_dis_mod2}
    \omega 
    &=
    i\gamma {\cal A}_0
    -iq^2 \frac{{\cal A}_2}{\gamma}
    +  {\cal O}(q^3)
    \ , \\
    \label{eq:sound_sindil}
    \omega
    &=
    \pm\frac{q}{\sqrt{d-1}}
    +iq^2  \frac{{\cal A}_2}{2\gamma}
    +   {\cal O}(q^3)\ .
\end{align}
where
\begin{align}
    {\cal A}_0
    &=
    -\frac{(d-1)\, s\, \frac{\partial^2P}{\partial\nu\partial T}
    }{\frac{\partial(s,n)}{\partial(T,\nu)}}
    \ , \qquad 
    {\cal A}_2 
    =
    \frac{(d-1) 
    \frac{\partial^2P}{\partial\nu^2} s
    -
    \frac{\partial(s,n)}{\partial(T,\nu)} T}{ (d-1)^2 s\,T
   \frac{\partial^2P}{\partial \nu \partial T}
   }
   \ .
\end{align}
The quadratic term in the dispersion relation of the dissipative mode \eqref{eq_dis_mod2} and that in the dispersion relations for the sound modes \eqref{eq:sound_sindil} are proportional but opposite in sign. Stability requires the leading imaginary parts to be positive, namely ${\cal A}_0\geq 0$ and ${\cal A}_2 \geq 0$. Assuming that all thermodynamic variables are positive is not enough to guarantee linear stability. In fact, we also need
\begin{eqnarray}
    \frac{\partial^2 P}{\partial T\partial\nu}
    <0
    \ , \qquad 
    \frac{\partial^2 P}{\partial\nu^2}
    \geq
    \frac{1}{s(d-1)} \left|\frac{\partial(s,n)}{\partial(T,\nu)}\right|\ .
\end{eqnarray}

In the conformal limit, $\gamma\to\infty$, the dissipative mode \eqref{eq_dis_mod2} features a leading term which diverges, while its subleading terms vanish. This corresponds to infinite damping, which effectively trivializes the dissipative mode. Note, however, that also in the non-conformal case, where $\gamma = 1$, there is no generic reason to expect ${\cal A}_0$ to be small. This implies that the extra mode falls outside the hydrodynamic regime and effectively decouples.

The propagating modes, instead, reduce to purely propagating sound for $\gamma\to\infty$, because the subleading terms vanish. The propagation velocity correctly coincides with the conformal speed of sound. In the non-conformal case, they are damped and relax upon considering a long enough time scale.

\subsection{With dilatonic condensate}
\label{sec:norm_with_dil}

The presence of a dilatonic condensate is accounted for through the dilaton Josephson relation and a non-trivial $h$, as described in Section \ref{sec:hydro}. In this case, the spectrum of hydrodynamic modes is given by two purely dissipative modes whose dispersion relation starts at order zero in momentum (``gapped") and a gapless sound branch. This latter is formed by two modified sound modes propagating at the conformal speed of sound, but featuring dissipation at subleading order. To avoid cumbersome expressions, we expand the leading coefficients of the purely dissipative modes with respect to $\gamma\to\infty$,
\begin{align}
    \label{eq:B0}
    \omega 
    &=
    \left[
    i {\cal B}_0 
    +
    {\cal O}\left(\gamma^{-1}\right)
    \right]
    +
    {\cal O}(q)
    \ , \\ 
    \label{eq:C0}
    \omega 
    &=
    \left[i \gamma\, {\cal C}_0 
    +
    {\cal O}\left(\gamma^0\right)
    \right] 
    +
    {\cal O}(q)\ ,
\end{align}
where
\begin{align}
    {\cal B}_0
    &\equiv 
    -
    \frac{ (d-2) P \left(s
    +
    \frac{\partial^2P}{\partial T^2} T\right) 
    -
    d\, s^2\,  T 
    }{\frac{\partial^2 P}{\partial\nu\partial T}\, s\, T}
    \ , \qquad
    {\cal C}_0 
    \equiv
    -(d-1) \, s\, \frac{\frac{\partial^2 P}{\partial\nu\partial T}}{\frac{\partial(s,n)}{\partial(T,\nu)}}
    \ .
\end{align}
The sound branch corresponds instead to 
\begin{align}
    \label{eq:sound_condil}
    \omega 
    &= 
    \pm \frac{q}{\sqrt{d-1}}
    +
    \frac{i\,q^2}{2\gamma} {\cal D}_2
    +
    {\cal O}(q^3)\ ,
\end{align}
with
\begin{align}
    {\cal D}_2
    &\equiv 
    -\frac{1}{(d-1)^2}
    \frac{d \frac{\partial^2P}{\partial\nu\partial T} \left[(d-2) P-s\, T\right]}
    {(d-2) P\left(s+\frac{\partial^2P}{\partial T^2} T\right)
    -d\, s^2\, T}
    \ .
    \label{eq:D2}
\end{align}

The subleading dissipative correction to the sound modes obtained in \eqref{eq:sound_condil}, ${\cal D}_2$, and that obtained in \eqref{eq:sound_sindil}, ${\cal A}_2$, are different. Thus the dissipation of the first sound is actually sensitive to the presence/absence of a dilatonic condensate. Nonetheless, this difference is eliminated in the conformal limit $\gamma\to\infty$. 

Let us now consider the implications of linear stability. These can be easily understood from the full expression for the gaps \eqref{eq:B0} and \eqref{eq:C0} where one does not expand in large $\gamma$ (we actually set $\gamma = 1$),
\begin{equation}
    \label{eq:opm}
    \omega_\pm^{(0)}
    = 
    \frac{i}{2 \, \left| \frac{\partial(T, \nu)}{\partial(s, n)} \right|\, T} \left( A\pm \sqrt{A^2+B}\right)\ ,
\end{equation}
with
\begin{align}
  A
  &=
  \frac{\partial^2P}{\partial T\partial\nu}\big[(d-2) P -  s \, T \left(2d-1\right) 
  \big]\ ,
  \\
  B 
  &=
  4 (d-1) \, \left| \frac{\partial(T, \nu)}{\partial(s, n)} \right| \, T \left[d \, s^2 \, T - (d-2) \, P \left( s + \frac{d^2P}{dT^2} \, T \right)\right] 
  \ .
\end{align}
As we have already observed in Subsection \ref{subsec:normal2}, assuming that all thermodynamic variables are positive is not enough to ensure linear stability. In fact, upon demanding ${\cal D}_2>0$ and $\text{Im}[\omega_\pm]>0$, one also obtains%
\footnote{The present discussion is not sensitive to the sign of the Jacobian $\frac{\partial(T, \nu)}{\partial(s, n)}$. For $d=2$, we seem to be forced to require a thermal state with zero entropy. Yet, we remind the reader that we actually know from the start that $d=2$ is not described by the present setup, see footnote \hyperlink{foot:d2}{4}.}
\begin{equation}
    \label{eq:bounds}
    \frac{(d-2) P}{(2d-1)T}<s<\frac{(d-2)P}{T} \ , \qquad
    \frac{\partial^2P}{\partial\nu\partial T}<0\ .
\end{equation}
We should notice that one might expect a correction to the  dissipative term \eqref{eq:D2} coming from second order hydrodynamics. With that caveat in mind, let us proceed with our analysis. Since $T$ is the only scale around, we have that $P = \alpha\, T^d$, where $\alpha$ is a numerical coefficient. Hence $s = \frac{\partial P}{\partial T} = d\, \alpha\, T^{d-1}$. Putting these into \eqref{eq:bounds}, we find that it is impossible to satisfy the bounds. Apparently, no stable system with dilatonic condensate is thermodynamically consistent, at least in the restricted setup where $T$ is the only thermodynamic scale.%
\footnote{
\hypertarget{foot:nuzero}{} \hspace{-8pt}
Actually $T$ is our only scale because we set $\nu$ to zero. Morally, we start with $T$ and $\nu$, so that the pressure is $P = T^d f(\frac{\nu}{T})$ where the function $f$ as well as its argument $\frac{\nu}{T}$ scale trivially, and then take the limit $\nu \to 0$. Thus, we are considering the large temperature limit, deeply in the normal phase. The fact that we cannot have the dilatonic condensate becomes therefore natural. There is also an analogy with the $U(1)$ superfluid, which generally requires a finite chemical potential. Relaxing $\nu=0$ is however a non-trivial generalization which we postpone to future investigation.}

In spite of the remark above, it is nevertheless technically interesting to study under which circumstances one might find a vanishing gap from \eqref{eq:opm}. Asking $\omega_-=0$ requires a fine-tuning of the specific susceptibilities 
\begin{equation}
    \frac{\partial^2P}{\partial\nu \partial T}=0
    \ , \qquad 
    \frac{\partial^2P}{\partial T^2}=\frac{d\, s^2}{(d-2)P}-\frac{s}{T}\ .
\end{equation}
Let us replace $s=\frac{\partial P}{\partial T}$ to get the equation
\begin{equation}
\frac{\partial^2P}{\partial T^2}=\left(\frac{\partial P}{\partial T}\right)^2\frac{d}{(d-2)P}-\frac{\partial P}{\partial T}\frac{1}{T} \ ,   
\end{equation}
that has a simple solution
\begin{equation}
    P= \Big[a(d-2)
    -
    2b \log T\Big]^{1-\frac d 2}\ .
\end{equation}
Avoidance of the logarithmic singularity requires one to set $b=0$, which leads to a trivialization of the thermodynamics.

\section{Final remarks}
\label{sec:remfut}

The fact that scale invariance implies conformal invariance in quantum field theory is a longstanding and oftentimes subtle field of research (see \cite{Nakayama:2013is} and references therein). The present analysis shows that the lack of conformal symmetry induces generic dissipative effects in scale-invariant hydrodynamics. This occurs regardless of the spacetime dimensionality, as long as it is greater than 2. Importantly, these dissipation effects arise already at ideal order and are eventually able to relax the hydrodynamic response completely, over long enough time scales.
It is therefore curious to observe that hydrodynamics, despite being just an effective framework, appears to be strongly biased towards conformality. This may be related to the fact that hydrodynamics is based on the same symmetries that govern the UV dynamics, and the generating functional from which it is derived, \eqref{eq:effac}, coincides with the low-energy leading term of the effective quantum action \cite{Son:2002zn}.

The conservation equation for scale symmetry \eqref{eq:WTI} equates the divergence of the Weyl current $I^\mu$ and the trace of the energy-momentum tensor. Thus, in order to be consistent with the derivative expansion, one needs the constitutive relation for $T^{\mu\nu}$ at first order in derivatives. Since we assume parity invariance, there are no possible first-order scalar quantities to be added to the effective action \eqref{eq:effac} (see \cite{Jensen:2012jh}). This implies that the constitutive relation for the energy-momentum tensor \eqref{eq:Tmunu} is valid up to the first order in derivatives. However, it would be interesting to study the effect of relaxing the assumption on parity. 

The considerations and conclusions deriving from the present analysis should be extendable to a non-relativistic context or, more generically, to fluids without boost symmetry \cite{deBoer:2017ing,deBoer:2017abi}. 
Another natural and interesting direction for future research consists in considering a finite $\nu$. This would also be needed to provide some clarification about the physical interpretation of $\nu$ itself. For our purposes, it is particularly interesting to understand whether a finite $\nu$ (and/or a finite chemical potential $\mu$ for an additional $U(1)$ symmetry) could rescue the dilatonic condensate from thermodynamic inconsistency (see comments in footnote \hyperlink{foot:nuzero}{9}). We expect that a thorough analysis might also give some insight on the condition $\bar n=0$ for $\bar\nu=0$ that we imposed in this work. Besides, considering an additional $U(1)$ symmetry would open a series of further questions regarding the physics of a scale invariant $U(1)$ superfluid and the possibly mixed dynamics of the hydrodynamic modes associated to the $U(1)$ Nambu-Goldstone and the dilaton, along similar lines as those described in the zero temperature analysis of \cite{Argurio:2020jcq}.

It is interesting to observe that the relation between scale and conformal invariance has already been addressed by means of low-energy effective approaches. Two notable examples are given by bottom-up holography \cite{Nakayama:2010wx,Flory:2025cyi} and the physics of slowly varying (although possibly strong) fields accounting for bosonic string theory at low-energy  \cite{Papadopoulos:2024uvi}. Another study based on coset techniques has been recently described in \cite{Itsios:2021eig, Itsios:2025fkx}.

\section*{Acknowledgements}

We would like to thank 
Riccardo Argurio, Francesco Bigazzi, Alessio Caddeo, Diego Correa, Aldo Cotrone, Carlos Hoyos, Antón Faedo, Roberto A. Lineros, Joaquín Otero Schroeder, Mario Queipo Silva, Alfonso Ramallo and Guille Silva for essential feedback and interesting discussions on various aspects of the present work. The work of EA is supported by the Severo Ochoa fellowship PA-23-BP22-170. The work of EA and DM has been partially supported by the AEI and the MCIU through the
Spanish grant PID2021-123021NB-I00. DN expresses his gratitude to the Wiener-Anspach Foundation for supporting his postdoctoral stay at the University of Cambridge. I.S.L would like to acknowledge support from the ICTP through the Associates Programme (2023-
2028).

\appendix 

\section{Details on the Ward-Takahashi identities}
\label{app:Ward-Takahashi}

The currents are defined in \eqref{eq:def_cur}. We can introduce a dilaton as a compensator field \cite{Komargodski:2011vj} to make the theory (manifestly) Weyl invariant ($\tau\to\tau+\sigma$):
\begin{align}
    \label{eq:varga}
    \delta \Gamma 
    &= 
    \int d^dx 
    \sqrt{-g} e^{d\tau}\Big[ 
    -\frac{1}{2} T^{\mu\nu} e^{(-w_T+2) \tau} \delta g_{\mu\nu}
    - I^\mu e^{(-w_I-w_\Omega) \tau} \delta \Omega_\mu
    \Big]\ .
\end{align}
We have indicated with $w_X$ the scaling dimension of the quantity $X$ and used $w_g = -2$,
\begin{align}
    &\left[g_{\mu\nu}\right] = w_g = -2
    \ , \quad 
    \left[I^\mu\right] = w_I
    \ , \quad 
    \left[\Omega_\mu\right] = w_\Omega\ .
\end{align}
The gauge transformations for the background fields are:
\begin{align}
    \label{eq:gau_Omega}
    \delta \Omega_\mu &= 
    e^{w_\Omega\tau} \partial_\mu \sigma 
    - w_\Omega \sigma \Omega_\mu 
    + \sigma  x^\nu \partial_\nu \Omega_\mu
    + \xi^\nu \partial_\nu \Omega_\mu
    + \Omega_\nu \partial_\mu \xi^\nu
    \\ \nonumber &\qquad 
    + b_\sigma (2x^\sigma x^\tau - g^{\sigma\tau}x^2)\partial_\tau \Omega_\mu
    + 2 x_\tau b_\sigma \left(w_\Omega g^{\sigma\tau} g_{\mu\nu}- J^{\sigma\tau}_{\mu\nu}\right) \Omega^\nu
    \ , \\
    \label{eq:gau_g}
    \delta g_{\mu\nu} &= 
    -w_g\sigma g_{\mu\nu}
    + \sigma  x^\alpha \partial_\alpha g_{\mu\nu}
    + \xi^\lambda \partial_\lambda g_{\mu\nu}
    + \partial_\mu \xi_\nu + \partial_\nu \xi_\mu
    \\ \nonumber &\qquad 
    + b_\sigma (2x^\sigma x^\tau - g^{\sigma\tau}x^2)\partial_\tau g_{\mu\nu}
    \\ \nonumber &\qquad 
    + 2 x_\tau b_\sigma \left(w_g g^{\sigma\tau} g_{\mu\alpha} g_{\nu\beta} - J^{\sigma\tau}_{\mu\alpha} g_{\nu\beta} - J^{\sigma\tau}_{\nu\beta} g_{\mu\alpha} \right) g^{\alpha\beta}
    \ ,
\end{align}
where we used the Lorentz generators
\begin{equation}
    \label{eq:gen_lor}
    J^{\lambda\sigma}_{\nu\mu}
    \equiv
    -i\left(
    \delta^\lambda_\nu \delta^\sigma_\mu 
    -
    \delta^\sigma_\nu \delta^\lambda_\mu 
    \right)\ .
\end{equation}
As far as diffeomorphisms are concerned, we have considered the Lie variations of the fields, without assuming flat backgrounds. The variations under special conformal transformations are discussed in \cite{DiFrancesco:1997nk,Ginsparg:1988ui,Jackiw:2011vz}.

Asking for invariance with respect to the gauge transformations of the background, we can derive the following Ward-Takahashi identities:
\begin{itemize} 
\item 
{\bfseries Weyl transformations:}
\begin{align}
    \nabla_\mu & \left(
    I^\mu 
    e^{(-w_I+d)\tau}
    \right) 
    -
    w_\Omega e^{(-w_I-w_\Omega+d)\tau} I^\mu \Omega_\mu
    + e^{(-w_I-w_\Omega+d)\tau} x^\alpha I^\mu \partial_\alpha \Omega_\mu
    \\ \nonumber &
    -
    w_g e^{(-w_T+d+2)\tau} T^{\mu\nu} g_{\mu\nu}
    +
    e^{(-w_T+d+2)\tau} x^\alpha T^{\mu\nu} \partial_\alpha g_{\mu\nu}
    = 
    0\ .
\end{align}
Setting $\Omega_\mu = 0$ and $g_{\mu\nu}=\eta_{\mu\nu}$ we get
\begin{equation}
    \label{eq:ska_ward}
    \nabla_\mu I^\mu 
    - (w_I-d) I^\mu \partial_\mu \tau
    -
    e^{(-w_T+2+w_I)\tau} T^{\mu}_{\ \mu}
    =
    0\ .
\end{equation}
This is the covariantization, also in the sense of Weyl symmetry \cite{Loganayagam:2008is,Bhattacharyya:2008mz}, of the standard scaling Ward-Takahashi identity
\begin{equation}
\label{eq:ska_war_tak}
    \partial_\mu D^\mu
    =
    \partial_\mu I^\mu + T^\mu_{\ \mu}  
    = 0\ ,
\end{equation}
where the scaling current is defined as 
\begin{equation}
    D^\mu 
    = 
    I^\mu + x_\nu T^{\mu\nu}\ ,
\end{equation}
and one assumes translation and Lorentz invariance, $\partial_\mu T^{\mu\nu}=0$ and $T^{[\mu\nu]}=0$.

\item 
{\bfseries Translations:}
\begin{align}
    & \nabla_\mu \left(T^{\mu\nu} e^{(-w_T+d+2)\tau}\right)
    - \frac{1}{2} T^{\alpha\beta} e^{(-w_T+d+2)\tau} \partial^\nu g_{\alpha\beta}
    \\ \nonumber & \qquad \qquad
    + \nabla_\mu \left(I^\mu e^{(-w_I-w_\Omega+d)\tau} \Omega^\nu\right)
    - I^\alpha e^{(-w_I-w_\Omega+d)\tau} \partial^\nu \Omega_\alpha
    = 0\ .
\end{align}
Taking $\Omega_\mu = 0$ and constant $g_{\mu\nu}$, it reduces to:
\begin{align}
    \label{eq:ward_trasla}
    \nabla_\mu T^{\mu\nu}
    - (w_T-d-2) T^{\mu\nu} \partial_\mu \tau
    = 0\ .
\end{align}
Eventually, taking into account $w_T = d+2$, we get
\begin{align}
    \nabla_\mu T^{\mu\nu}
    = 0\ .
\end{align}
\end{itemize}

\subsection{Conformal Ward-Takahashi identity}
\label{app:conformal}

The variation of the background fields under special conformal transformations can be read from \eqref{eq:gau_Omega} and \eqref{eq:gau_g} by setting $\sigma=0$ and $\xi^\mu=0$. The conformal Ward-Takahashi identity when assuming a vanishing background Weyl connection, $\Omega_\mu = 0$, reads
\begin{equation}
    T^{\mu\nu} x_\tau b_{\sigma}\left(2 g^{\sigma\tau} g_{\mu\alpha} g_{\nu\beta} 
    + 
     J^{\sigma\tau}_{\mu\alpha} g_{\nu\beta} 
    + J^{\sigma\tau}_{\nu\beta} g_{\mu\alpha} \right) g^{\alpha\beta} = 0\ .
\end{equation}
Given that $J^{\lambda\sigma}_{(\nu\mu)} = 0$, it eventually leads to the tracelessness condition for the energy-momentum tensor
\begin{equation}
    \label{eq_conf}
    T^{\mu}_{\ \mu} = 0\ .
\end{equation}

\section{Scale invariant but not conformal examples}
\label{sec:sio}

We examine some explicit field-theory examples that exhibit scale invariance without conformal invariance, focusing mainly on the scaling Ward-Takahashi identity.

\subsection{Free Maxwell theory in $d\neq 4$}

Maxwell theory in $d\geq 3$ with $d\neq 4$ spacetime dimensions is scale but not conformal invariant \cite{Jackiw:2011vz,El_Showk_2011,Nakayama:2013is}. From the action 
\begin{equation}
    \label{eq::Maxwell}
    S=\int d^dx\, \sqrt{-g}\, {\cal L}
    \ , \qquad 
    {\cal L} = -\frac{1}{4}F_{\mu\nu}F^{\mu\nu} \ ,
\end{equation}
one gets the energy-momentum tensor
\begin{equation}
    \label{eq::energymomentum::Maxwell}
    T^{\mu\nu}=\frac{-2}{\sqrt{-g}}\frac{\delta {\cal L}}{\delta g_{\mu\nu}}=-\frac{g^{\mu\nu}}{4}F^2+F^{\alpha\nu}F_{\alpha}^{\,\,\mu}\ .
\end{equation}
Its trace reads
\begin{equation}
    \label{eq::traceenergymomentum::Maxwell}
    T^{\mu}_{\ \mu}
    =
    \frac{4-d}{4}F^2\approx\frac{4-d}{2}\partial_{\mu}\left(F^{\mu\nu}A_{\nu}\right)
    =
    \partial_{\mu}k^{\mu}\ ,
\end{equation}
where the equality $\approx$ holds on-shell. We have defined the virial current
\begin{equation}
    \label{eq:virial_Maxwell}
    k^{\mu} \equiv \frac{4-d}{2} F^{\mu\nu}A_\nu\ .
\end{equation}
Notice that $T^{\mu}_{\ \mu}$ is non-zero in $d\neq 4$ and it cannot be improved to be vanishing, which would require it to be a total derivative. The analysis of scale vs conformal invariance in $d=4$ led to some interesting developments, see for instance \cite{Dymarsky:2013pqa, Naseh:2016maw}. 

In order to encounter the scaling current $J^{\mu}_D$, we follow the Noether procedure considering $\Delta A_{\mu}=\sigma w_A A_{\mu}$ and $\Delta x^{\nu}=-\sigma x^{\nu}$,
\begin{eqnarray}
    J^{\mu}_D
    &=&
    \frac{\delta \mathcal{L}}{\delta  \partial_{\mu}A_{\nu}}\frac{\delta \Delta A_{\nu}}{\delta \sigma}
    +
    \left[\delta^{\mu}_{\nu}\mathcal{L}-\frac{\delta \mathcal{L}}{\delta \partial_{\mu}A_{\alpha}}\partial_{\nu}A_{\alpha} \right]\frac{\delta\Delta x^{\nu}}{\delta\sigma}  \\
    &\approx&
    (1-w_A) F^{\mu\nu}A_{\nu}
    -
    x^{\nu} T^{\mu}_{\ \nu}
    -
    \partial_{\alpha}\left( x^{\nu}F^{\mu\alpha}A_{\nu}\right) \nonumber 
    \ ,
\end{eqnarray}
where $T^\mu_{\ \nu}$ is the Belinfante energy momentum tensor \eqref{eq::energymomentum::Maxwell}, connected to the canonical energy-momentum tensor through the last, total-derivative term.

Unlike in Section \ref{sec:WT}, here we are not implementing scale symmetry \emph{à la} Weyl, in fact coordinates and derivatives are transforming; thus, from Maxwell action we have $w_A=\frac{d-2}{2}$.%
\footnote{
This is not in contradiction with the result $w_A=w_\Omega=0$ described in Appendix \ref{app:closing_the_algebra}. Introducing suitably scaling factors as $\partial_\mu A_\nu = e^{\tau} \tilde \partial_\mu (e^{w_A\tau} \tilde A_\nu)$, one could implement the scaling transformation internally as $\tau\to \tau+\sigma$. The field $\tilde A_\nu$ and the coordinates would thus be scale invariant and we could follow Noether's procedure as usual by promoting the scaling parameter $\sigma$ to be local.
} Therefore, we get
\begin{equation}
    \label{eq::scalingcurrent}
    J^{\mu}_D
    =
    \frac{4-d}{2}F^{\mu\nu}A_{\nu}-x^{\nu}T^{\mu}_{\ \nu}\ ,
\end{equation}
where we dropped an improvement term. Using \eqref{eq::traceenergymomentum::Maxwell} we get that the scaling current $J^{\mu}_D$ is conserved, 
\begin{equation}
    \partial_{\mu}J^{\mu}_D
    =
    \frac{4-d}{2}\partial_{\mu}\left(F^{\mu\nu}A_{\nu}\right)
    -
    T^{\mu}_{\ \mu} 
    = 
    0\ ,
\end{equation}
having used that $T^{\mu\nu}$ is locally conserved. Note that we are getting an explicit realization of the Ward-Takahashi identity \eqref{eq:scaling_conservation_Weyl} upon identifying the Weyl current as
\begin{eqnarray}
    I^\mu 
    =
    \frac{4-d}{2} F^{\mu\nu}A_{\nu}\ ,
\end{eqnarray}
which in fact coincides with the virial current \eqref{eq:virial_Maxwell}. For comments regarding the lack of gauge invariance for $J_D^\mu$ we refer to \cite{El_Showk_2011}.

\subsection{Covariant elastic-like theories}

We examine the class of non-unitary models proposed in \cite{Riva_2005}, whose action is
\begin{equation}
    S=\int d^dx\, \sqrt{-g}\, {\cal L}
    \ , \qquad 
    \label{eq:lag_vec2}
    {\cal L} = \frac{1}{4}u_{\mu\nu}u^{\mu\nu}-\frac{\alpha}{2}(\partial_{\mu}u^{\mu})^2
    \ ,
\end{equation}
where the strain tensor is defined as $u_{\mu\nu}=\partial_{\mu}u_{\nu}+\partial_{\nu}u_{\mu}$ and $\alpha$ is a coupling constant. The equations of motion are
\begin{equation}
\label{eq::eqom::massless2}
    \partial_{\mu}\left( u^{\mu\nu}-\alpha \, g^{\mu\nu}\partial_{\rho}u^{\rho}\right)=0\ ,
\end{equation}
while the Belinfante-Hilbert energy-momentum tensor is
\begin{align}
    T^{\mu\nu}
    &=\frac{-2}{\sqrt{-g}}\frac{\delta {\cal L}}{\delta g_{\mu\nu}}
    \\ \nonumber &=
    \frac{1}{4} g^{\mu\nu}\left[u_{\alpha\beta}u^{\alpha\beta}-2\alpha(\partial_{\rho}u^{\rho})^2
    \right]
    -u^{\nu}_{\,\,\beta}u^{\mu\beta}+2\alpha \partial_{\rho}u^{\rho}\partial^{(\mu}u^{\nu)}
    \ .
\end{align}

Considering the scaling transformations $\Delta u_{\mu}=\sigma w_u u_{\mu}$ and $\Delta x^{\nu}=-\sigma x^{\nu}$ and following the Noether procedure, we get the scaling current $J^{\mu}_D$
\begin{eqnarray}
\label{eq::scalingcurrent::massless2}
    J^{\mu}_D&=&\frac{\delta \mathcal{L}}{\delta \partial_{\mu}u_{\nu}}\frac{\delta \Delta u_{\nu}}{\delta \sigma}
    +
    \left[\delta^{\mu}_{\nu}\mathcal{L}-\frac{\delta \mathcal{L}}{\delta \partial_{\mu}u_{\alpha}}\partial_{\nu}u_{\alpha} \right]\frac{\partial \Delta x^{\nu} }{\partial\sigma}
    \\ \nonumber
    &=& 
    (1+w_u) \left(u^{\mu\nu} - \alpha g^{\mu\nu} \partial_\rho u^\rho \right)
    u_\nu
    -
    T^{\mu}_{\ \nu}x^{\nu}
    -\partial_\alpha L^{\mu\alpha}
    \ ,
\end{eqnarray}
with 
\begin{eqnarray}
    L^{\mu\alpha}
    \equiv 
    \left(u^{\mu\alpha}-\alpha\,  g^{\mu\alpha}\partial_\rho u^\rho\right)u^\nu x_\nu\ .
\end{eqnarray}
Fixing $w_u=\frac{d-2}{2}$ and taking the divergence of \eqref{eq::scalingcurrent::massless2} yields 
\begin{equation}
    \label{eq::divergence::massless2}
    \partial_{\mu}J^{\mu}_D
    =\frac{d}{2}\,
    \partial_{\mu}
    \left(u^{\mu\nu} - \alpha g^{\mu\nu} \partial_\rho u^\rho \right) u_\nu
    -
    T^{\mu}_{\ \mu}
    -\partial_\mu \partial_\alpha\, L^{\mu\alpha} \ .
\end{equation}
We can proceed in analogy with \cite{Polchinski:1987dy,DiFrancesco:1997nk} and improve the energy-momentum tensor as follows:
\begin{eqnarray}
\label{eq:imp_ten}
    \tilde T^{\mu\nu}
    = 
    T^{\mu\nu}
    +
    \frac{1}{2} \partial_\lambda \partial_\rho X^{\lambda\rho\mu\nu}\ ,
\end{eqnarray}
with
\begin{align}
    X^{\lambda\rho\mu\nu}
    &=
    \frac{2}{d-2}\Big[
    g^{\lambda\rho}L^{\mu\nu}
    - g^{\lambda\mu}L^{\rho\nu}
    - g^{\rho\nu}L^{\lambda\mu}
    + g^{\mu\nu}L^{\lambda\rho}
    \\ \nonumber &\qquad 
    - \frac{1}{d-1}\left(
    g^{\lambda\rho}g^{\mu\nu}
    -
    g^{\lambda\mu}g^{\rho\nu}
    \right) L^\alpha_{\ \alpha}
    \Big]\ .
\end{align}
In fact, given the symmetry properties of the tensor $X$, we have that the improvement term $\frac{1}{2}\partial_\lambda\partial_\rho X^{\lambda\rho\mu\nu}$ is transverse and symmetric in the $\mu, \nu$ indexes; thus
\begin{equation}
    \partial_\mu \partial_\lambda \partial_\rho X^{\lambda\rho\mu\nu}
    =
    0\ .
\end{equation}
Moreover, we explicitly get  
\begin{equation}
    \frac{1}{2} g_{\mu\nu} \partial_\lambda \partial_\rho X^{\lambda\rho\mu\nu}
    =
    \partial_\alpha \partial_\beta L^{\alpha\beta}\ ,
\end{equation}
so that the trace of \eqref{eq:imp_ten} gives
\begin{eqnarray}
    \tilde T^\mu_{\ \mu}
    = 
    T^\mu_{\ \mu}
    +
    \partial_\alpha \partial_\beta L^{\alpha\beta}\ .
\end{eqnarray}
The scaling conservation equation then takes the form 
\begin{equation}
    \partial_{\mu}\tilde{J}^{\mu}_D
    =
    \partial_{\mu} I^{\mu}-\tilde T^{\mu}_{\ \mu}=0\ .
\end{equation}
The Weyl current coincides with the virial current.

\section{Closing the gauge transformation algebra}
\label{app:closing_the_algebra}

We reconsider the gauge transformations \eqref{eq:gauge_transformations}, this time allowing for non-trivial scaling dimensions for the background connection $\Omega_\mu$,
\begin{align}
    \delta \Omega_\mu 
    =
    {\cal L}_\xi \Omega_\mu 
    + w_\Omega\, \sigma\, \Omega_\mu 
    + \partial_\mu \sigma
    \ .
\end{align}
The gauge transformation for the metric is the same as in \eqref{eq:gauge_transformations}. We study the closure of the algebra of the gauge transformations. Namely, we require that the commutator of two transformations is still a transformation of the same kind:
\begin{equation}
    \label{eq:algebra}
    \left[\delta_1,\delta_2\right]
    =
    \delta_{[12]}\ .
\end{equation}
Explicitly, we encounter
\begin{align}
    \delta_{[12]} g^{\mu\nu}
    &=
    \left(\xi_2^\rho \partial_\rho \xi_1^\lambda - \xi_1^\rho \partial_\rho \xi_2^\lambda \right) \partial_\lambda g^{\mu\nu}
    -
    g^{\mu\lambda}\partial_\lambda \left(\xi_2^\rho\partial_\rho\xi_1^\nu - \xi_1^\rho\partial_\rho\xi_2^\nu\right)
    \\ \nonumber & \quad 
    -
    g^{\nu\lambda}\partial_\lambda \left(\xi_2^\rho\partial_\rho\xi_1^\mu - \xi_1^\rho\partial_\rho\xi_2^\mu\right)
    \ , \\ \label{eq:Omega_closure}
    \delta_{[12]} \Omega_\mu
    &=
    \left(\xi_2^\rho\partial_\rho\xi_1^\lambda - \xi_1^\rho\partial_\rho\xi_2^\lambda\right) \partial_\lambda \Omega_\mu
    +
    \Omega_\lambda \partial_\mu \left(\xi_2^\rho\partial_\rho\xi_1^\lambda - \xi_1^\rho\partial_\rho\xi_2^\lambda\right)
    \\ \nonumber &\quad 
    + 
    w_\Omega \left({\cal L}_{\xi_2} \sigma_1 - {\cal L}_{\xi_1} \sigma_2\right) \Omega_\mu
    +
    \partial_\mu\left({\cal L}_{\xi_2}\sigma_1 - {\cal L}_{\xi_1}\sigma_2\right)
    \\ \nonumber & \quad 
    + 
    w_\Omega \left(\sigma_2\partial_\mu\sigma_1 - \sigma_1\partial_\mu\sigma_2\right)
    \ .
\end{align}
The algebra closes only if $w_\Omega=0$ and we have 
\begin{align}
    \xi^\lambda_{[12]}
    &=
    \xi_2^\rho\partial_\rho\xi_1^\lambda - \xi_1^\rho\partial_\rho\xi_2^\lambda
    \ ,\\ 
    \sigma_{[12]}
    &=
    {\cal L}_{\xi_2} \sigma_1 - {\cal L}_{\xi_1} \sigma_2
    \ .
\end{align}

The extra terms that prevent the closure for $w_\Omega\neq 0$ are not central extensions. In fact, this would entail extending \eqref{eq:algebra} as follows
\begin{equation}
    \left[\delta_{1},\delta_{2}\right]
    =
    \delta_{[12]}
    +
    f_{(1,2)}\, \mathbb 1\ ,
\end{equation}
where $f_{(1,2)}$ is a function of the parameters of the gauge transformations 1 and 2 and it is antisymmetric for $1\leftrightarrow 2$. Referring for instance to the transformations for $\Omega_\mu$ in \eqref{eq:Omega_closure}, we identify 
\begin{equation}
    \label{eq:alpha_central_extension}
    f_{(1,2)} = -\sigma_2 \partial_\mu \sigma_1 + \sigma_1 \partial_\mu \sigma_2\ .
\end{equation}
Imposing that the extended algebra still satisfies the Jacobi identity translates into the 2-cocycle condition \cite{Lekeu:2021flo}
\begin{equation}
    f_{([12],3)} + f_{([31],2)} + f_{([23],1)} = 0\ ,    
\end{equation}
which is not satisfied by \eqref{eq:alpha_central_extension}. This agrees with the fact that, for spacetime dimension $d>2$, any central extension of the Poincaré algebra plus scaling is trivial as it can be reabsorbed by a suitable redefinition of the algebra generators \cite{Nakayama:2023xzu}. Including the special conformal transformation into the study of the closure of the gauged algebra will not change the discussion about the necessity to set $w_\Omega = 0$.

\section{Hydrodynamic conservation equations}
\label{app:hydrocon}

From the constitutive relation of the Weyl current, \eqref{eq:weycur}, considered at equilibrium (order zero in the fluctuations), we have%
\footnote{
Later on, we will enforce $n = 0$ as a consequence of $\nu = 0$ as established in \eqref{eq:Teq}; namely we assume that the Weyl density vanishes at equilibrium when its corresponding chemical potential is zero. 
}
\begin{equation}
    I^\mu = \delta^\mu_0 e^{d\tau}\, n
    \ .
\end{equation}
At linear order, the Weyl current reads
\begin{align}
    \nonumber
    \delta I^\mu &=
    d\, \delta\tau \, \delta^\mu_0 e^{d\tau} \, n
    +
    \delta^\mu_0 e^{d\tau} \delta n
    +
    \delta^\mu_0 e^{d\tau} \delta (h^2 \nu)
    +
    \delta^\mu_i e^{d\tau} \left(n\delta v^i+h^2\delta V^i\right)
    \\ \nonumber &
    =
    d\, \delta\tau \, \delta^\mu_0  e^{d\tau} \, n
    +
    \delta^\mu_0 e^{d\tau} \left(
    \frac{\partial^2 P}{\partial T\partial\nu}\delta T
    +
    \frac{\partial^2 P}{\partial\nu^2}\delta \nu
    \right)
    \\  &\qquad+
    \delta^\mu_i e^{d\tau} \left(n\delta v^i+h^2\delta V^i \right)\ ,
\end{align}
so that its conservation corresponds to
\begin{align}
    \nonumber
    \partial_\mu \delta I^\mu &=
    d\, e^{d\tau} \, n  \partial_0 \delta\tau
    +
    e^{d\tau}\frac{\partial^2 P}{\partial T\partial\nu}\partial_0\delta T
    +
    e^{d\tau}\frac{\partial^2 P}{\partial\nu^2}\partial_0\delta \nu
    \\ \label{eq:Weyldiv}
    &\qquad+
    e^{d\tau}n\partial_i\delta v^i
    +
    e^{d\tau}h^2\partial_i\delta V^i
    \ ,
\end{align}

To complete the hydrodynamic system of equations, we need to consider the energy-momentum tensor and its conservation at linear order in the fluctuations. The philosophy is the same as that followed for the Weyl current, nonetheless the technical development is slightly more involved. First, at zero-order the energy-momentum tensor is 
\begin{align}
    T^{\mu\nu} &=
    e^{d\tau}
    \left( 
    \delta^\mu_0 \delta^\nu_0 \, s T 
    -
    \eta^{\mu\nu} P
    \right)
    \ .
\end{align}
Its trace is given by
\begin{align}
    T^\mu_{\ \mu} &=
    e^{d\tau}
    \left(
    s T-d P 
    \right)
    \ .
\end{align}

Let us study the linear fluctuation of the various terms entering the expression of the energy-momentum individually:
\begin{align}
    &\delta\left(-\hat g^{\mu\nu}P\right) 
    =
    -\eta^{\mu\nu}\left[
    -2P\,\delta\tau
    +
    s\delta T 
    + n  \delta\nu\,\right]\ ,
    \\ 
    &\delta\left(s T u^\mu u ^\nu\right) 
    =
    \delta^\mu_0 \delta^\nu_0 \left[
    \left(\frac{\partial^2 P}{\partial T^2}\delta T
    +\frac{\partial^2P}{\partial\nu\partial T}\delta\nu\right)T  
    + s \delta T \right]
    + s T \left( 
    \delta^\mu_i \delta^\nu_0
     + \delta^\nu_i \delta^\mu_0
    \right) \delta v^i\ ,
    \\
    &\delta \left(n\nu u^\mu u^\nu\right)
    =
    \delta^\mu_0 \delta^\nu_0\, 
    n\,\delta\nu
    \ , \\
    &\delta \left(h^2 V^\mu V^\nu\right)
    =
    0
    \ ,
\end{align}
where we have already used that, at equilibrium, $u^i=V^i=0$ together with $\nu=0$. Therefore, for the fluctuation of the energy density we get
\begin{align}
    \delta T^{00}
    &= 
    e^{d\tau}\Bigg\{
    \delta \tau
    \left[
    d \, s T
    -(d-2)P 
    \right] 
    + \delta T\, \mathcal D s
    +\delta \nu\, \mathcal D n
    \Bigg\}\ .
\end{align}
The fluctuation for the momentum density is
\begin{equation}
    \delta T^{0i} 
    = 
    \delta T^{i0} 
    =
    e^{d\tau}
    sT \delta v^i
    \ .
\end{equation}
Eventually, the fluctuation of the purely spatial components is given by
\begin{equation}
    \delta T^{ij}
    = 
    e^{d\tau}\delta^{ij} \left[(d-2)P\,  \delta\tau 
    +
    s\delta T 
    + n \delta\nu\right]\ .
\end{equation}
Collecting the various terms together, energy conservation leads to
\begin{align}
    \partial_\mu \delta T^{\mu 0} 
    =
    e^{d\tau}&\Bigg\{
    \partial_0 \delta \tau
    \left[
    d\, s T
    -(d-2) P\right] 
    + 
    \partial_0\delta T \, \mathcal Ds
    +
    \partial_0\delta\nu
    \, \mathcal D n
    + sT \partial_i\delta v^i
    \Bigg\}\ ,
\end{align}
while momentum conservation corresponds to
\begin{align}
    \partial_\mu \delta T^{\mu i} &=
    e^{d\tau} \Bigg\{(d-2)\, P\, \partial^i \delta\tau 
    +
    sT \partial_0\delta v^i
    + s\partial^i\delta T 
    + 
    n  \partial^i\delta\nu
    \Bigg\}\ .
\end{align}

The last explicit piece of computation we need in order to be able to write the system of hydrodynamic equations regards the trace of the energy-momentum tensor,
\begin{align}
    \label{eq:antrace}
    &\delta \left(T^{\alpha\beta}\eta_{\alpha\beta}\right) 
    = 
    e^{d\tau} \Bigg\{\delta \tau\,d\left(s T
    -d P
    \right)  
    + \delta T \left[ 
    \mathcal D
    -(d-1) 
    \right]s 
    +
    \delta\nu\left[ 
    \mathcal D
    -(d-1)
    \right] n
    \Bigg\} 
    \ ,
\end{align}
where $\mathcal D$ is the differential operator introduced in \eqref{eq:D_operator}.

\bibliographystyle{utphys}
\bibliography{bib}

\end{document}